\documentclass[twocolumn,aps,prl,showpacs,amsmath,superscriptaddress]{revtex4}
\usepackage{amssymb}
\usepackage{mathrsfs}
\usepackage{graphicx}
\usepackage{float}

\begin{document}

\title{Quantum criticality in disordered bosonic optical lattices }

\author{Xiaoming Cai}
\affiliation{Beijing National Laboratory for Condensed Matter Physics, Institute of
Physics, Chinese Academy of Sciences, Beijing 100190, China}
\author{Shu Chen}
\email{schen@aphy.iphy.ac.cn} \affiliation{Beijing National Laboratory for Condensed
Matter Physics, Institute of Physics, Chinese Academy of Sciences, Beijing 100190,
China}
\author{Yupeng Wang}
\affiliation{Beijing National Laboratory for Condensed Matter Physics, Institute of
Physics, Chinese Academy of Sciences, Beijing 100190, China}
\date{ \today}

\begin{abstract}
Using the exact Bose-Fermi mapping, we study universal properties of ground-state
density distributions and finite-temperature quantum critical behavior of
one-dimensional hard-core bosons in trapped incommensurate optical lattices. Through
the analysis of universal scaling relations in the quantum critical regime, we
demonstrate that the superfluid to Bose glass transition and the general phase diagram
of disordered hard-core bosons can be uniquely determined from finite-temperature
density distributions of the trapped disordered system.
\end{abstract}


\maketitle

{\it Introduction.-} In recent years ultracold atomic systems have proven to be
powerful quantum simulators for investigating various challenging many-body physics. In
comparison with traditional condensed matter systems, both the potential and
interaction of trapped atomic systems can be experimentally implemented in a controlled
way, which leads to experimental breakthroughs in the study of fundamental model
systems, including the realization of the Tonks-Girardeau (TG) gas
\cite{Paredes,Kinoshita,Haller}, and the observation of Anderson localization in
disordered bosonic optical lattices \cite{Roati,Billy,Pasienski,Deissler}.
The interplay of disorder and interactions for bosons has often been studied in the
context of Bose-Hubbard model with disordered or incommensurate potentials. Theoretical
studies have indicated that a superfluid to Bose-glass (BG) transition occurs for
increasing disorder \cite{Fisher,Giamarchi}. Despite intensive studies
\cite{Fisher,Giamarchi,numerical,Fontanesi}, no exact solutions for disordered
interacting bosons are known except that in the TG limit the one-dimensional (1D)
disordered Bose-Hubbard models are exactly solvable via a Bose-Fermi mapping
\cite{Egger,Cai1,Cai2}.

So far most of theoretical studies on 1D disordered bosonic gases focused on zero
temperature cases, as finite-temperature phase transitions are generally absent in 1D
systems. However, considering realistic experiments are always carried on at finite
temperature for trapped systems, it is important to understand how to unambiguously
determine the zero temperature phase diagram from the knowledge of finite temperature
density profiles of trapped gases \cite{Zhou,Guan,Mueller}. In the presence of a
harmonic trap, Anderson plateaus are found in the average density profile of the
hard-core bosons in the incommensurate optical lattice \cite{Cai2}. The Anderson
plateau can be viewed as the signature of Anderson localized states with superfluid to
BG transitions taking place at edges of the plateau.  At finite temperatures, Anderson
plateaus are smeared out by thermal fluctuations and thus no sharp boundaries for
superfluid and BG phases could be detected. In order to understand zero temperature
phase transitions, we must consider the so-called quantum critical regime where some
universal scaling relations govern the finite-temperature physics. In principle, one
can extract the zero-temperature phase diagram of the disordered Bose system from these
universal relations. Unfortunately, it is very hard to calculate these universal
functions due to lack of reliable general techniques for disordered many-body problems.
Since quantum and thermal fluctuations strongly couple together in the quantum critical
regime, highly precise calculation of the finite-temperature data is needed in order to
extract correct zero-temperature phase diagram. In this work, by using the exact
numerical method based on the Bose-Fermi mapping, we can calculate finite-temperature
properties of disordered hard-core bosons exactly, and thus demonstrate that the
zero-temperature phase diagram can be uniquely determined from finite-temperature
density distributions of the trapped gas.


To make progress in unveiling the quantum criticality in disordered Bose systems
quantitatively, we study a disordered 1D TG gas with the effect of disorder being
mimicked by an incommensurate potential \cite{Roati,Deissler}, which is described by
\begin{equation}
\label{eqn1} H=-t\sum_i(\hat{b}^\dagger_i \hat{b}_{i+1}+ \mathrm{H.c.})+\sum_iV_i
\hat{n}_i,
\end{equation}
where $\hat{b}^\dagger_i$ ($\hat{b}_i$) is the creation (annihilation) operator of the
boson fulfilling the hard-core constraints, {\it i.e.,} the on-site anticommutation
$(\{ \hat{b}_i, \hat{b}^\dagger_i\}=1)$ and $[ \hat{b}_i, \hat{b}^\dagger_j]=0$ for
$i\neq j$, and $\hat{n}_i=\hat{b}^\dagger_i \hat{b}_i$. The hopping amplitude $t$ is
set to be the unit of the energy $(t=1)$, and $V_i$ is given by
\begin{equation}
\label{eqn2} V_i=V_I\mathrm{cos}(\alpha2\pi i+\delta)+V_H(i-i_0)^2.
\end{equation}
Here $V_I$ is the strength of incommensurate potentia with $\alpha$ being an irrational
number
and $\delta$ an arbitrary phase which is chosen to be zero for convenience, and $V_H$
is the strength of the additional harmonic trap with $i_0$ being the position of trap
center.
In the TG limit, the Hamiltonian (\ref{eqn1}) can be mapped to a noninteracting
spinless Fermi model, which allows us to calculate even the finite-temperature
properties by using the exact numerical method \cite{Rigol1,Rigol_T}.

\begin{figure}[tbp]
\includegraphics[scale=0.6]{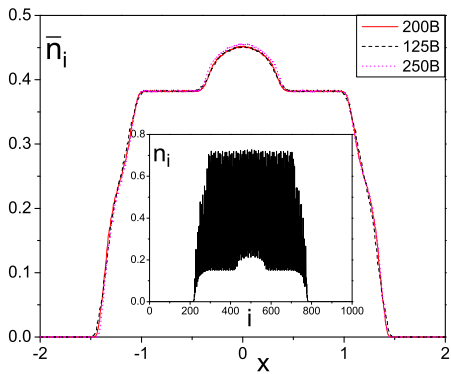}
\caption{(Color online) Local average density profiles relative to the scaled position
for different systems with same characteristic density($\widetilde{\rho}=1$). The three
systems are: 200 bosons, $V_H=2.5\times10^{-5}$; 125 bosons, $V_H=6.4\times10^{-5}$;
and 250 bosons, $V_H=1.6\times10^{-5}$ with 1000 sites, $V_I=1$ and
$\alpha=(\sqrt{5}-1)/2$. Insert: corresponding density profile for the system with 200
bosons.} \label{Fig1}
\end{figure}

{\it Universal properties of zero-temperature density distribution.-} In order to
characterize universal features of lattice systems in harmonic traps, we make use of
the length scale
$\zeta=(V_H/t)^{-1/2}$ and the characteristic density $\widetilde{\rho}=N/\zeta$
introduced in Ref.\cite{Rigol2}. Without the incommensurate potential the system has
been studied by Rigol and Muramatsu in detail \cite{Rigol1}, and they found that there
is a critical characteristic density in the system ($\widetilde{\rho}_c\sim2.6-2.7$).
For low characteristic density( $\widetilde{\rho}<\widetilde{\rho}_c$), the whole
system is in a uniform phase (superfluid phase) at zero temperature, and for
$\widetilde{\rho}>\widetilde{\rho}_c$ two different phases coexist with a Mott
insulating plateau in the middle of the trap
surrounded by superfluid phases on two sides. In the presence of incommensurate
potential \cite{Cai2}, the density profiles basically still have the arc shape for weak
$V_I$, but there are a lot of drastic oscillations induced by the incommensurate
potential (see the insert picture in Fig.\ref{Fig1}).
In order to reduce the drastic oscillations in density profiles, we define the local
average density $\overline{n}_i=\sum_{j=-M}^M n_{i+j}/(2M+1)$, where $2M+1$ is the
length to count the local average density with $M\ll L$ and
$n_i=\langle\Psi^G_{\mathrm{HCB}}| \hat{n}_i |\Psi^G_{\mathrm{HCB}}\rangle$ with
$\Psi^G_{\mathrm{HCB}}$ the ground state of the Hamiltonian (\ref{eqn1}). When $V_I
\rightarrow 0$, the local average density profile is almost the same as the density
profile. As $V_I$ increases, Anderson plateaus appear at shoulders of the arc, then
become wider and wider but the height of the plateau does not change with $V_I$. For
sites in the Anderson plateau the one particle density matrices
$\rho_{ij}=\langle\Psi^G_{\mathrm{HCB}}| \hat{b}^\dagger_i
\hat{b}_j|\Psi^G_{\mathrm{HCB}}\rangle$ have an exponential-law decay which is the
character of the particle in the BG phase. While for sites outside the plateau,
$\rho_{ij}$ exhibit a power-law decay which is the character of the particle in the
superfluid phase.

\begin{figure}[tbp]
\includegraphics[scale=0.6]{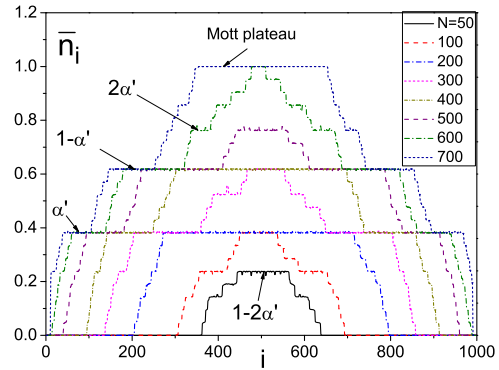}
\caption{(Color online) Local average density profiles for systems with different
particle number (characteristic density). The other parameters of system are 1000
sites, $V_I=2.4$, $\alpha=(\sqrt{5}-1)/2$ and $V_H=2.5\times10^{-5}$.} \label{Fig2}
\end{figure}

Now we discuss the universal properties of Anderson plateaus.
In Fig.\ref{Fig1} we show local average density profiles for three systems with the
same characteristic density relative to the scaled position $x=(i-i_0)/\zeta$. They are
almost identical except of tiny differences caused by the setting of $M=10$ for all
three systems with different length scale $\zeta$. So for systems with different
particle number and strength of the harmonic trap, Anderson plateaus are the same so
long as these systems have the same characteristic density. From now on, we fix the
strength of harmonic trap and change particle numbers to achieve different
characteristic densities. By doing this we don't need to scale the position for
comparison among different systems. As shown in Fig.\ref{Fig2}, we plot local average
density profiles for systems with fixed $V_H$ and different particle numbers (different
$\widetilde{\rho}$). Here we choose $V_I>2$ to generate as many Anderson plateaus as
possible in local average density profiles. Due to the incommensurate potential taking
the form of $\mathrm{cos}(2\pi \alpha i)$, for any $\alpha\in (-\infty,\infty)$ one can
always choose a corresponding $\alpha'\in [0,0.5)$ which generates the same
incommensurate potential \cite{Cai1}. After counting on the locations of Anderson
plateaus, we find that the height of Anderson plateaus is decided by $\alpha'$ with
values $\alpha', 1-\alpha', 2\alpha', 1-2\alpha', 2(1-\alpha'), 1-2(1-\alpha'),
4\alpha', 1-4\alpha',...$, if the values are in the range of $(0,1)$. The Anderson
plateaus with $\bar{n}_i=\alpha', 1-\alpha', 2\alpha', 1-2\alpha'$ are the most
important as they are wide enough and easy to appear for $V_I<2$. So heights of
Anderson plateaus are totally decided by $\alpha'$, while the width and the existence
of the Anderson plateau are associated with $V_I$ and the characteristic density
($\widetilde{\rho}$). For system with high characteristic density, except of the
existence of Anderson plateaus, there is a Mott plateau in the trap center
characterized by $n_i=1$.

\begin{figure}[tbp]
\includegraphics[scale=0.6]{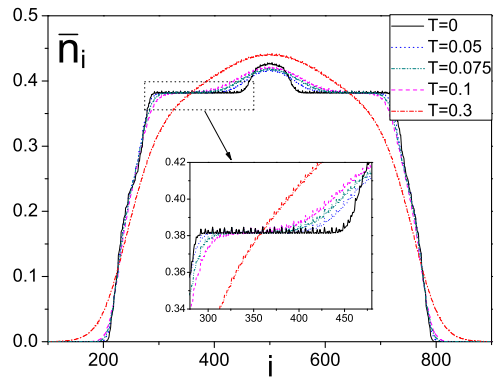}
\caption{(Color online) Local average density profiles for systems with different
temperatures. Insert: enlargement of the Anderson-plateau area. The system is with 1000
sites, 200 bosons, $V_I=1.3$, $\alpha=(\sqrt{5}-1)/2$ and $V_H=2.5\times10^{-5}$.}
\label{Fig3}
\end{figure}

{\it Finite-temperature density distribution and quantum criticality.-} The
finite-temperature density distribution can be calculated by
\begin{equation}
n_i(T)=\frac{1}{Z}\sum_{n=1}^{N_s} e^{-E_n/k_B T} \langle\Psi^n_{\mathrm{HCB}}|
\hat{b}^\dagger_i \hat{b}_i|\Psi^n_{\mathrm{HCB}}\rangle,
\end{equation}
where $N_s=L!/(L-N)!N!$ is the number of states, $E_n$ is the energy of eigenstate
$\Psi^n_{\mathrm{HCB}}$, and $Z=\sum_{n=1}^{N_s} e^{-E_n/k_B T}$ is the canonical
partition function. By using the exact numerical method in Ref.\cite{Rigol_T}, we can
calculate the finite-temperature properties of the hard-core bosons in incommensurate
lattice very efficiently. In Fig.\ref{Fig3}, we display the local average density
profiles for systems with different temperatures. The obvious plateau at zero
temperature tends to vanish with the increase in the temperature. Consequently, the
sharp boundary for superfluid and Anderson localized phases is destroyed by temperature
fluctuations. If the temperature is low and located in the quantum critical regime, the
density distributions for different temperatures should fulfil some universal scaling
laws around the quantum phase transition point.

Before going to the analysis on the quantum criticality, we would like to introduce
briefly the general theory for the quantum criticality. Given the equation of state
$n=n(T,\mu)$ for a system with density $n$, temperature $T$ and chemical potential
$\mu$, then near the zero-temperature phase transition point $\mu=\mu_c$, it was shown
\cite{Zhou,Fisher} that when the dimensionality of the system is below a critical
dimension, the following universal relation exists:
\begin{equation}
\label{eqn3} n(\mu,T)-n_r(\mu,T)=T^{\tfrac{d}{z}+1-\tfrac{1}{\nu
z}}\Omega(\tfrac{\mu-\mu_c}{T^{1/\nu z}}),
\end{equation}
with $n_r(\mu,T)$ being the regular part of the density, $d$ the dimensionality of the
system, $\nu$ the correlation length exponent, and z the dynamical exponent.
$\Omega(x)$ is a universal function which describes the singular part of the density
near criticality. If $n_r(\mu,T)$ in the above equation is known for one or more phases
(such as vacuum and Mott insulator), Eq.(\ref{eqn3}) would be very useful for
determining the quantum phase transition point. After scaling the density with the $T$
term, we get $A(\mu,T)=T^{-\tfrac{d}{z}-1+\tfrac{1}{\nu z}}(n(\mu,T)-n_r(\mu,T))$. If
we plot $A(\mu,T)$ versus $\mu$ for a system at different temperatures, then curves
with different temperatures will intersect at the same point $\mu=\mu_c$. With this
scaling method, we can detect critical points of phase transitions at zero temperature
from none-zero-temperature density profiles. Once $\mu_c$ is determined, we can scale
the chemical potential $\widetilde{\mu}=(\mu-\mu_c)/T^{\tfrac{1}{\nu z}}$. By plotting
$A(\mu,T)$ versus $\widetilde{\mu}$ for a system at different temperatures, all curves
collapse into a single one (the universal function $\Omega(x)$). It was shown that $d =
1$, $z=2$ and $\nu=1/2$ for both the vacuum-superfluid and superfluid-Mott phase
transition of hard-core bosons \cite{Zhou}.

\begin{figure}[tbp]
\includegraphics[scale=0.8]{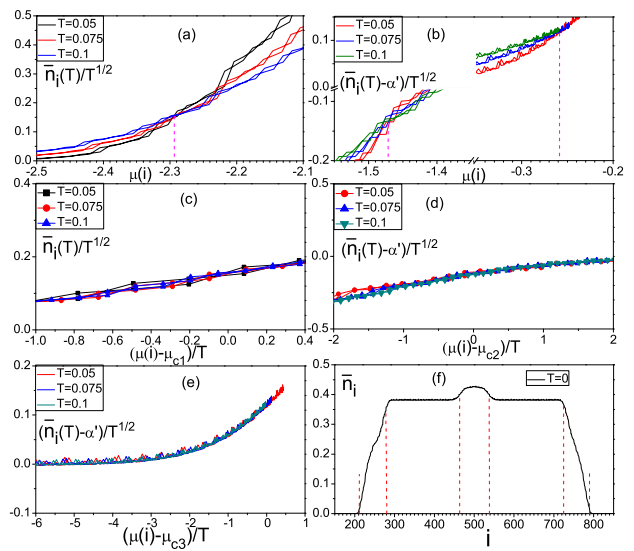}
\caption{(Color online) (a,b): Scaled local average density profiles vs $\mu(i)$ at
different temperatures. (c,d,e): Scaled local average density profiles vs scaled
chemical potential at different temperatures for $\mu_{c1}$(c), $\mu_{c2}$(d),
$\mu_{c3}$(e). (f): The local average density profile of the system at zero temperature
together with the critical point $i_c$ got from the scaling method. The system is with
1000 sites, 200 bosons, $V_I=1.3$, $\alpha=(\sqrt{5}-1)/2$ and $V_H=2.5\times10^{-5}$.}
\label{Fig4}
\end{figure}

For hard-core bosons in the 1D incommensurate lattice within a harmonic trap, there may
exist vacuum-superfluid, superfluid-BG, and even the superfluid-Mott insulator phase
transition as the position $i$ changing from sides to the center. If we replace the
chemical potential $\mu$ in Eq.(\ref{eqn3}) by $\mu(i)=\mu-V_H(i)$, where $V_H(i)$ is
the harmonic trap potential and $\mu$ is determined by $\sum_i n(\mu(i),T) = N$, then
$\mu(i)$ will drive the transition from one phase to the other phase. Now we explore
the quantum criticality of the disordered Bose system by using the local average
density distribution. We find that there exist similar scaling laws described by
Eq.(\ref{eqn3}) but with the density distribution in Eq.(\ref{eqn3}) being replaced by
the local average density distribution. To give concrete examples, we consider a system
with 200 hard-core bosons, $V_I=1.3$, $\alpha=(\sqrt{5}-1)/2$ and $V_H=2.5\times
10^{-5}$, for which there is no Mott insulating plateau in the local average density
profile as the characteristic density $\widetilde{\rho}=1<\widetilde{\rho}_c$. In
Fig.\ref{Fig4}a, we plot scaled densities $\bar{n}_i (T)/T^{1/2}$ for different
temperatures around the edge of the density profile, where $\bar{n}_i (T) \equiv
\bar{n}(\mu_i,T)$. Different curves intersect at the point $\mu_{c1}(i)=-2.293$, which
is just the zero-temperature vacuum-superfluid transition point and occurs at sites
$i_{c1}=788.2$ and $212.8$ via $\mu(i)=\mu-V_H(i)$.

From Fig.\ref{Fig3} we know that Anderson plateaus appear at $\bar{n}_i =
\alpha'=0.38197$. By taking the regular part $\bar{n}_r=\alpha'$, in Fig.\ref{Fig4}b we
plot scaled densities $(\bar{n}_i (T)-\alpha')/T^{1/2}$ versus $\mu(i)$ around the
regime of Anderson plateau for different temperatures. It is clear that the different
curves intersect at two points, which indicates that the superfluid-BG phase
transitions occur at $\mu_{c2}(i)=-1.472$ and $\mu_{c3}(i)=-0.258$£¬or equivalently at
$i_{c2}=724$ and $277$, and at $i_{c3}=537.6$ and $463.4$. To display the universal
scaling functions, we plot scaled local average densities at different temperatures
against scaled chemical potentials around critical points $\mu_{c1}$, $\mu_{c2}$,
$\mu_{c3}$ in Fig.\ref{Fig4}c, d, e. We observe that curves for different temperatures
indeed collapse onto a single curve after scaling, except that there are oscillations
on the curves caused by the incommensurate lattice. In Fig.\ref{Fig4}f we show the
zero-temperature local average density profile in contrast to critical points
determined by the scaling method and marked by dash lines.
\begin{figure}[tbp]
\includegraphics[scale=0.6]{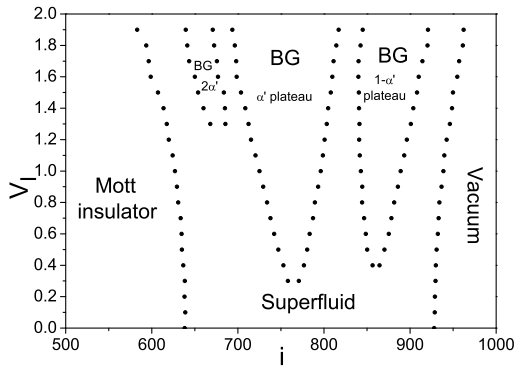}
\caption{ The right half of the phase diagram at zero temperature in $V_I$-position
plane of the system with 1000 sites, 600 bosons, $\alpha=(\sqrt{5}-1)/2$ and
$V_H=2.5\times10^{-5}$.} \label{Fig5}
\end{figure}

Through the analysis of quantum criticality, we can map out the zero temperature phase
diagram from the nonzero-temperature local average density distributions. As an
example, we show the phase diagram of the system with 1000 sites, 600 bosons,
$\alpha=(\sqrt{5}-1)/2$ and $V_H=2.5\times10^{-5}$ in the $V_I$-position plane, which
has the characteristic density $\widetilde{\rho}=3>\widetilde{\rho}_c$. As shown in
Fig.{\ref{Fig5}}, without the incommensurate potential the system has a structure with
the Mott insulator in the center surrounding by the superfluid phase on two sides. As
$V_I$ increases, Anderson plateaus appear in the local average density profiles and the
particles in them are in BG phase. With further increase in $V_I$, more plateaus appear
and become wider. Furthermore, the size of Mott insulator decreases but the system size
grows slowly as $V_I$ increases.

In summary, using the Bose-Fermi mapping, we study the universal properties and quantum
criticality of one-dimensional hard-core bosons on incommensurate optical lattices
within harmonic traps. By calculating finite-temperature density distributions exactly,
we unveil universal scaling relations of local average density distributions in the
quantum critical regime. The zero temperature phase diagram is then determined from
nonzero temperature local average density profiles with help of the analysis of quantum
criticality.
\begin{acknowledgments}
We thank Xi-Wen Guan and Qi Zhou for helpful discussions. This work has been supported
by NSF of China under Grants No.10821403 and No.10974234, 973 grant and National
Program for Basic Research of MOST.
\end{acknowledgments}

\end{document}